# Regulating Ride-Sourcing Markets: Can Minimum Wage Regulation Protect Drivers Without Disrupting the Market?

**Farnoud Ghasemi**[1,*], **Arjan de Ruijter**[2], **Rafal Kucharski**[1], **Oded Cats**[2]

[1]Faculty of Mathematics and Computer Science, Jagiellonian University, Krakow 30-348, Poland
[2]Department of Transport & Planning, Delft University of Technology, Delft 2628 CN, The Netherlands
[*]farnoud.ghasemi@doctoral.uj.edu.pl

## ABSTRACT

Ride-sourcing platforms such as Uber and Lyft are prime examples of the gig economy, recruiting drivers as independent contractors, thereby avoiding legal and fiscal obligations. Although platforms offer flexibility in choosing work shifts and areas, many drivers experience low income and poor working conditions, leading to widespread strikes and protests. Minimum wage regulation is adopted to improve drivers' welfare. However, the impacts of this regulation on drivers as well as on travelers and platforms, remain largely unknown. While ride-sourcing platforms do not disclose the relevant data, state-of-the-art models fail to explain the effects of minimum wage regulation on market dynamics. In this study, we assess the effectiveness and implications of minimum wage regulation in ride-sourcing markets while simulating the detailed dynamics of ride-sourcing markets under varying regulation intensities, both with and without the so-called platform lockout strategy. Our findings reveal that minimum wage regulation impacts substantially drivers income, and may lead to higher fares for travelers and threaten platforms' survival. When platforms adopt a lockout strategy, their profitability significantly improves and drivers earn more, although many others lose their jobs, and service level for travelers consequently declines.

## Introduction

The gig economy revolution, centered around digital two-sided platforms, has transformed markets across diverse sectors, from mobility (Uber) and delivery (Just Eat) to freelancing (Upwork) and household services (TaskRabbit). These platforms connect independent workers (freelancers) with consumers seeking on-demand goods and services, by implementing charges. The seamless interaction facilitated by platforms between supply and demand leads to network effects, where the value created by the platform on one side increases as more users join, either on the same side (same-side network effects) or opposite side (cross-side network effects), driving substantial expansion of the platform [1]. The growing gig economy is not only transforming industries but also reshaping one of the most fundamental aspects of the economy and society: the labor market. Flexibility, autonomy, and low entry barriers are key factors attracting gig workers, whether they seek full-time job or supplementary income [2]. However, there are serious ethical concerns surrounding gig work. In particular, the classification of gig workers as non-employees, which results in weak social and legal protections. Indeed, platforms seek to avoid legal and fiscal obligations by identifying gig workers as self-employed independent contractors, thereby significantly reducing operating costs. Furthermore, precarious and often menial nature of gig work contributes to the exploitation of workers [3]. Thus, regulations, such as guaranteed minimum wage, play a crucial role in addressing these concerns, ensuring that the rapid growth of the gig economy does not come at the expense of workers' rights and protections.

Ride-sourcing (ride-hailing) platforms like Uber, Lyft, and Didi constitute a significant share of the gig economy labor market, with gig workers serving as drivers. In the United States, for example, ride-sourcing drivers represent 23% of all gig workers, the largest share among the other sectors [4]. While the primary reason drivers are drawn to these platforms is the flexibility they offer in choosing working shifts and areas, many drivers complain about their low income [5]. Platforms frequently advertise potential earnings of $25–$35 per hour; however, actual net hourly wages for drivers are estimated to range between $5.72 and $10.46 per hour [6]. The disparity between drivers' expected and actual income, along with poor working conditions, has led to numerous protests, strikes, and lawsuits against platforms worldwide, drawing attentions to drivers' rights [7,8]. The courts in the United Kingdom and the Netherlands ruled that ride-sourcing drivers must be classified as employees rather than independent contractors, ensuring that drivers are entitled to the rights of traditional employees [9,10]. In several countries, including the United States [11,12], China [13], and Brazil [14], regulations have been implemented to guarantee that drivers receive a minimum wage.

There are several factors underlying the low driver income in ride-sourcing markets. One major factor is demand fluctuation, which varies by time of day, location, and seasonality, often leading to reduced driver income during periods of low demand [15,16]. However, even during times of high demand, when drivers experience minimal idle time, they may still face low earnings. This can be attributed to platform strategies, such as setting low per-kilometer fares to attract more passengers and imposing high commission rates to increase platform revenue [17]. Furthermore, an oversupply of drivers increases competition for rides and diminishes individual earnings, highlighting the importance of maintaining a supply-demand balance in the market [18]. In addition to low income, drivers encounter high volatility in their earning as it depends on unpredictable demand patterns and platform strategies [15].

While minimum wage regulation for ride-sourcing markets aim to address drivers low income and prevent oversupply, its outcomes vary significantly across cities, not only for ride-sourcing drivers but also for the platforms and travelers. The reasons underlying the differing impacts of minimum wage regulation on ride-sourcing market dynamics remain largely unknown. For instance, Proposition 22, passed in California in 2020, aimed to provide certain benefits to ride-hailing drivers, including a guaranteed earnings rate of 120% of the local minimum wage. However, after its implementation, drivers' net income remained significantly below the state's minimum wage [19]. New York City implemented a minimum wage standard for ride-sourcing drivers, requiring platforms to pay drivers a minimum of $17.22 per hour (after expenses), which roughly corresponds to the $15 hourly state minimum wage in 2019 when accounting for payroll taxes [20]. While the regulation improved drivers' income, unintended consequences emerged. The higher guaranteed wages attracted more drivers to the market, increasing competition for rides and reducing the number of trips completed by each driver. Platforms like Uber and Lyft responded by increasing trip fares and implementing the so-called lockout strategy in New York City, which restricts access to the platform for certain drivers during specific times, to avoid excessive costs. While the lockout strategy reduced working opportunities, particularly for part-time drivers, the fare increases led to a reduction in demand [21]. To this end, the inherent complexity of ride-sourcing markets, driven by the non-linear interactions among key stakeholders—platforms, drivers, and travelers—impedes the effective implementation of minimum wage regulation. This highlights the need for a nuanced understanding of the multifaceted impacts of minimum wage regulation on the complex dynamics of ride-sourcing markets to enable its effective implementation.

Ride-sourcing platforms operate with a significant lack of transparency, restricting access to critical data on platform dynamics relevant to minimum wage regulation. At the same time, state-of-the-art models used to study minimum wage regulation fail to provide a comprehensive understanding of this regulation and its disruptive impact on ride-sourcing markets. Prior studies mainly associate the effectiveness of minimum wage regulation with the implementation approach, the responses of ride-sourcing platforms, and local labor market conditions shaping drivers' expectations [3,22,23]. Li et al. [24] examine the regulation of ride-sourcing markets using a queuing equilibrium model, suggesting that enforcing a minimum wage for drivers could improve the social welfare at the expense of the platform. In another study, Gurvich et al. [22] employ a capacity management model and argue that implementing a minimum wage forces the platform to restrict the number of drivers active on the platform during specific time periods, which could be detrimental to social good. Zhang and Nie [25] emphasize the importance of inter-platform competition in studying minimum wage regulation. Using a game-theoretic approach, they argue that in competitive ride-sourcing markets, a minimum wage mitigates the negative consequences of self-destructive pricing wars, leading to improvements in traveler and driver surplus, as well as platform profits. Parrott and Reich [26] utilize administrative data collected from platforms to simulate the effects of the minimum wage regulation in New York City. They estimate that the regulation leads to only a minor increase in traveler waiting time, and a minimal fare adjustment would suffice to compensate platforms for the driver pay increase. Sun et al. [27] employ the System Dynamics method to examine platform growth under regulatory policies, reproducing the dynamics of the ride-sourcing market. While this is an important aspect that the aforementioned approaches are unable to capture, their model fails to generate the network effects, fundamental elements of market dynamics. Conversely, agent-based modeling has shown to be a suitable method for examining regulations and their far reaching impact, capturing the behavior of drivers and travelers, as well as platform strategies, which give rise to network effect [17]. Using an agent-based model, de Ruijter et al. [18] find that platforms thrive on socio-economic inequality within ride-sourcing markets. They suggest that this outcome arises from the combination of cheap labor and time-sensitive ride-sourcing users, reinforced by the network effects inherent to these markets. These studies are either equilibrium-based (non-dynamic), fail to reproduce key market dynamics, mainly network effects, or disregard inter-platform competition by assuming a monopoly market structure. Surprisingly, platforms' lockout strategy has been largely overlooked in the existing literature, despite its widespread use in practice and significant effect on market dynamics in competitive market structures. Therefore, an adequate understanding of minimum wage regulation in complex ride-sourcing markets is missing, and its impact on the dynamic nature of these markets remains largely unexplored. While the variation in driver income and service levels for travelers resulting from the introduction of minimum wage regulations by regulators and lockout strategies by platforms remains an open question, it remains unknown how this regulation affects platforms' viability.

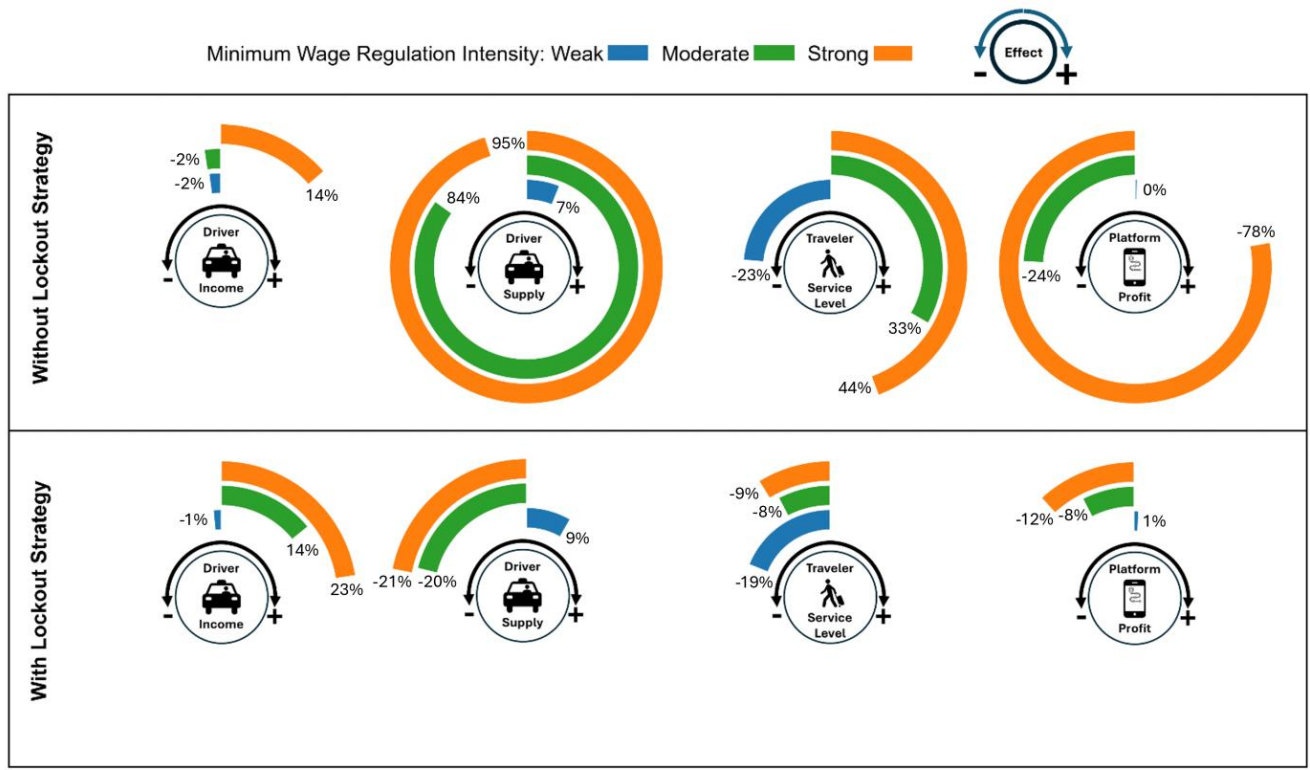

**Figure 1.** The impact of three level of minimum wage regulation relative to no regulation condition on the ride-sourcing market evaluated through: driver income, driver supply, traveler service level, and platform profitability. We explore the regulation at three levels relative to the average driver's income expectation: (i) weak—20% below expectation, (ii) moderate—at expectation, and (iii) strong—20% above expectation. Our experimental results suggest that without platforms' lockout (upper row), driver supply—which reflects the availability of job opportunities—increases across all three levels of regulation, whereas driver income rises only under strong regulatory conditions. While service level for traveler declines only under weak regulation, platform profitability consistently decreases, and severely so under strong regulation. With platforms' lockout (lower row), we observe an inverse relationship between changes in driver income and driver supply. Driver income increases further—with a notable rise even under moderate regulation—while driver supply declines, except for under weak regulatory conditions. While Traveler's level of service declines under all regulation levels, platforms mitigate potential negative impact to a large extent by applying the lockout strategy

To address this gap, we examine the impact of minimum wage regulation at varying levels on drivers, travelers, and platforms, both in the presence and the absence of platforms' lockout strategy (see Figure. 1). To this end, we extend the MoMaS agent-based framework [28] with an inter-platform competition model to reproduce the detailed dynamics of ride-sourcing markets under minimum wage regulation. In this model, driver and traveler agents make utility-based mode choices between platforms and alternative transport modes on a daily basis, while platforms compete over shared pools of supply and demand through trip fare adjustments. The model provides a realistic representation of non-linear interactions among drivers, travelers, and platforms which give rise to positive and negative network effects in the market. The proposed model does not only offer a deeper understanding of minimum wage regulation in competitive ride-sourcing markets but also equips policymakers and regulators with actionable insights to design and implement this regulation effectively.

## Application and Results

We apply the ride-sourcing simulation model, detailed in methodology section, to the case of Amsterdam, the Netherlands, with a pool of 2000 travelers and 200 drivers—sufficient to reproduce network effects [1]. The simulation is run over a period of 2000 days on the detailed road network of the city retrieved from OpenStreetMap, with each simulated day representing a 4-hour time window from 08:00 to 12:00. Each traveler is assigned a trip drawn from the real-world Albatross trip set [29] and makes a daily mode choice between two ride-sourcing platforms and public transport. The quality of public transport service is based on GTFS data for Amsterdam. Each driver makes a daily decision between working for a ride-sourcing platform and pursuing an alternative occupation with a fixed reservation wage of €12 per hour. In this context, the reservation wage refers to the minimum wage rate a driver could earn in an alternative occupation. The vehicle speed is assumed to be constant across the road network at 36 kilometer per hour. Traveler and driver agents learn platforms' utility via three utility components, with the following weights: $\beta_i^e = 0.70$ for the weight of experienced utility, $\beta_i^{wom} = 0.20$ for the weight of word-of-mouth utility, and $\beta_i^m = 0.10$ for the weight of marketing utility [28]. Both platforms strategically adjust their trip fares over the course of competition with a turnover interval of 50 days. Each applies a fixed 20% commission rate and incurs a daily operational expense of €500 throughout the simulation.

Such a modeling framework enables a realistic investigation of the impact of minimum wage regulation on the dynamics of ride-sourcing markets, allowing for the evaluation of key performance indicators for platforms, travelers, and drivers. In the baseline scenario, no regulation is applied. In the other scenarios, we consider three levels of minimum wage regulation intensity—weak, moderate, and strong. Additionally, each of these regulation levels is simulated under two conditions: with and without the platform lockout strategy. This results in a total of seven experimental scenarios. The intensity of the minimum wage regulation is defined relative to the driver's reservation wage. Specifically, weak regulation is set at 20% below the reservation wage, corresponding to €9.60 per hour; moderate regulation is aligned with the reservation wage, set at €12.00 per hour; and strong regulation is fixed at 20% above the reservation wage, amounting to €14.40 per hour. Under these regulations, the minimum wage serves as a guaranteed hourly income for drivers. If a driver's earnings fall below the threshold, the platform must cover the shortfall, ensuring all active drivers earn at least the regulated amount—regardless of trip demand or operational efficiency. The lockout strategy enables platforms to restrict drivers' access to the platform on certain days, thereby avoiding excessive wage compensation costs under the regulation.

The results are structured into three subsections to systematically unpack the effects of minimum wage regulation and platform behavior on market outcomes. First, we examine the influence of minimum wage regulation in isolation—disregarding any strategic platform response—in order to identify its baseline impact on market dynamics and stakeholder welfare. Next, we incorporate the lockout strategy employed by platforms, analyzing how this platform strategy adaptation interacts with wage regulations to reshape the market dynamics. Finally, we investigate how both regulatory and strategic factors influence the distribution of driver income and the level of service experienced by travelers, drawing on the full set of experimental scenarios examined in the previous subsections. Notably, our analysis focuses on overall market-level dynamics rather than platform-specific outcomes, as the platforms showed broadly similar performance. Consequently, the reported indicators represent aggregated results across the two platforms.

## Trip Fare Outcomes of Inter-platform Competition

The equilibrium fare levels reached by both platforms at the end of the pricing game vary across regulatory scenarios. The equilibrium fare under the no-regulation scenario is 1.4. In the absence of platform lockout strategies, the equilibrium fares are 1.2 under weak regulation, 1.2 under moderate regulation, and 1.6 under strong regulation. In the presence of platform lockout strategies—applicable only under regulatory scenarios—the equilibrium fares are 1.2 under weak regulation, 1.6 under moderate regulation, and 1.6 under strong regulation.

It is important to note that trip fares influence both the supply and demand sides of the market. While travelers might benefit from lower fares, drivers experience short-term gains from higher fares. However, in the long term, feedback loops within the system may reverse these initial effects. For example, lower trip fares can reduce driver earnings, leading to a decline in the number of active drivers, which may subsequently degrade service quality and reduce traveler demand. Conversely, higher trip fares can discourage traveler demand, eventually leading to a decline in driver participation due to reduced ride volume and income opportunities. This dynamic creates a fundamental trade-off for platforms: fare levels must balance affordability for travelers and sustainability for drivers to ensure platform viability in the market.

Under the weak regulation scenario, platform trip fares stabilize at €1.20 per kilometer with both lockout and non-lockout conditions. This fare level is slightly below the equilibrium trip fare of €1.40 per kilometer observed in the no-regulation scenario. By setting a lower fare, platforms effectively slow the growth of supply in the market (see Figures 3 and 5), which in turn leads to a significantly reduced daily wage subsidy requirement. The same pricing behavior is observed under moderate regulation without the lockout condition—not only in terms of the resulting fare level, but also in the underlying platform strategy to limit driver supply and minimize subsidy payments. In contrast, under moderate regulation with lockout, platforms converge to a higher equilibrium fare of €1.60 per kilometer. This is because the number of active drivers is already reduced by the lockout, prompting platforms to extract greater profit per trip by raising fares. Under the strong regulation, platforms stabilize their trip fare at €1.60 per kilometer, regardless of whether the lockout strategy is applied. In this context, platforms face significant wage subsidy imposed by minimum wage regulations. Setting a lower fare does not help platform viability, as it results in both reduced revenue. Consequently, platforms opt for a higher fare to offset these costs to some extent and maintain financial sustainability.

## Impact of Minimum Wage Regulation

We first investigate the impact of varying levels of minimum wage regulation on the ride-sourcing market in the absence of the platform lockout strategy. Figure 2 illustrates the two-sided market evolution, capturing supply growth (a) and demand growth (c), alongside key performance indicators: driver income (b) as a proxy for driver satisfaction, and traveler waiting time (d) as a measure of service quality. Considering the number of drivers and travelers across all scenarios, supply and demand growth is initially slow until reaching a critical mass, after which it accelerates due to positive cross-side network effects and eventually stabilizes as the market is exploited.

On the supply side, the driver income under the no regulation scenario stabilizes at approximately €12.50 per hour, slightly above the driver reservation wage (RW) of €12.00 per hour. Weak regulation appears to have a negligible effect on both the number of drivers and their income. Interestingly, the number of drivers under moderate regulation increases significantly, despite no improvement in driver income. This is because drivers do not experience any negative outcomes, as their income is always equal to or greater than their reservation wage. Unsurprisingly, drivers are most satisfied under strong regulation, earning €14.20 per hour on the average, as income above their reservation wage is guaranteed. Consequently, drivers adopt the platforms more quickly under this regulation, achieving the highest market share compared to the other scenarios.

On the demand side, with no regulation the number of travelers exhibits balanced growth relative to the number of drivers. Consequently, the average traveler waiting time stabilizes at approximately 3.5 minutes. Under weak regulation, the market experiences a slight undersupply, as the number of drivers remains nearly unchanged while the number of travelers increases compared to the no regulation scenario. This results in a deterioration in the level of service for travelers, with their average waiting time reaching to 4.40 minutes. In contrast, moderate regulation leads to a slight oversupply in the market, reducing the average waiting time to 2.40 minutes. Surprisingly, under strong regulation, the market exhibits the lowest demand level despite offering travelers a waiting time as low as 2.00 minutes on the average. This is attributed to the fact that travelers' willingness to use the platforms depends not only on waiting time but also on trip fare. Trip fares resulting from platform competition stabilize at a lower level of €1.20 per kilometer under both weak and moderate minimum wage regulations, compared to €1.40 per kilometer in the absence of regulation. In contrast, under strong regulation, fares converge to a higher level of €1.60 per kilometer relative to the no regulation scenario.

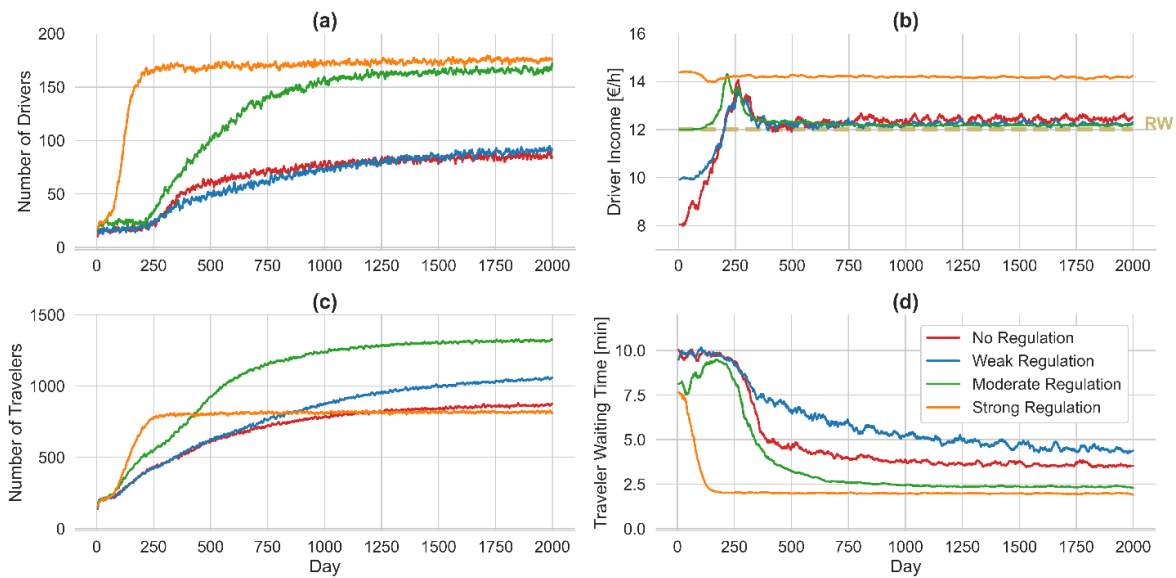

**Figure 2.** Ride-sourcing supply and demand performance under varying minimum wage regulations without platform lockout strategy. Four scenarios are illustrated based on the presence and intensity of regulation: no regulation (red), weak regulation (blue), moderate regulation (green), and strong regulation (orange). On the supply side (top row), the number of active drivers and their average income are shown. On the demand side (bottom row), the number of travelers and their average waiting time are depicted.

Minimum wage regulations affects not only the behavior and welfare of drivers and travelers but also have significant implications for the viability of ride-sourcing platforms. Figure 3 presents the average performance of two platforms under minimum wage regulations across key indicators. The daily revenue trends are highly correlated with the number of travelers, with strong regulation resulting in the lowest gains and moderate regulation leading to the highest gains for platforms. Since no regulation is applied in the baseline scenario, the daily subsidy—i.e., the platform's compensation to ensure all drivers earn at least the determined minimum wage—remains zero throughout the simulation. Under weak regulation, this subsidy remains slightly above zero, indicating that most active drivers earn more than the €9.60 per hour minimum wage solely from rides. On the other hand, under moderate regulation, platforms stabilize their daily subsidies at approximately €1100, whereas under strong regulation, they incur significantly higher subsidies of around €2600 per day.

Under weak regulation, platforms exhibit a trend similar to the no-regulation scenario, achieving long-term profitability due to the minimal subsidy burden imposed. Under moderate regulation, platforms experience positive profitability growth in the short term, primarily due to the limited number of active drivers. However, as the number of drivers increases after day 250—when platforms begin incurring substantial subsidy costs—profitability declines, eventually stabilizing at an average daily loss of approximately €620. The large number of participating drivers under strong regulation in the short term results in substantial subsidy payments by platforms, leading to escalating losses that reach approximately €2400 per day by the end of the simulation. Accumulated capital captures the platform's cumulative financial trajectory by aggregating daily gains and losses over time, offering a long-term measure of sustainability under varying regulatory conditions. Under weak regulation, platforms gradually accumulate capital, indicating long-term financial sustainability; in contrast, under moderate and strong regulations, accumulated capital consistently declines into increasingly negative territory, signaling potential bankruptcy in the absence of external financial support.

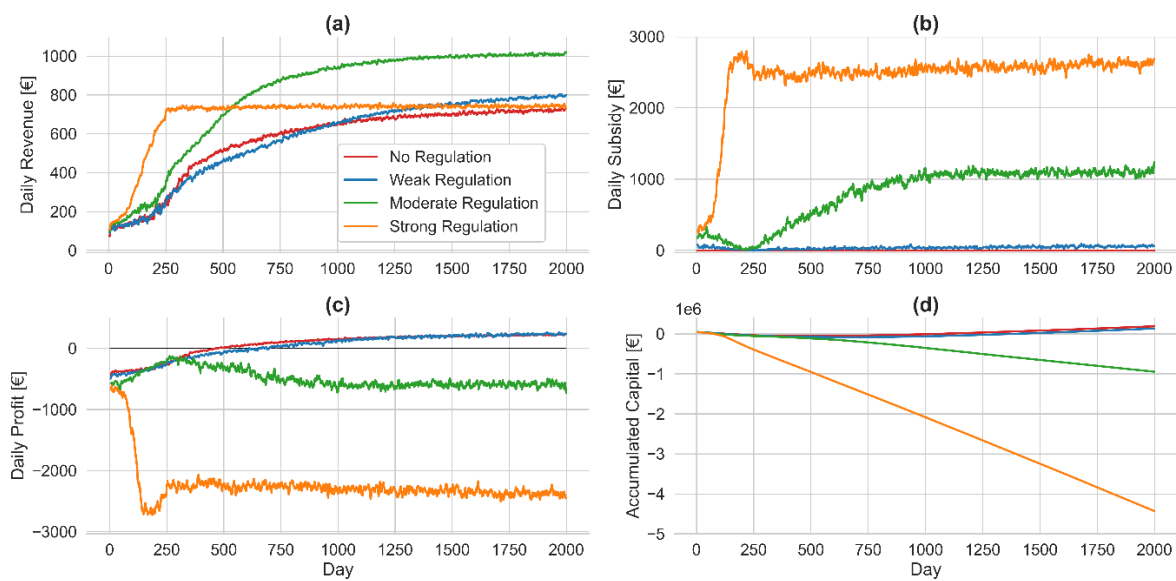

**Figure 3.** Average performance of two platforms under varying minimum wage regulations without lockout strategy. Four scenarios are illustrated based on the presence and intensity of regulation: no regulation (red), weak regulation (blue), moderate regulation (green), and strong regulation (orange). Daily platform revenue is derived from a fixed 20% commission on trip fares. Daily subsidy represents the total compensation paid by platforms to ensure driver earnings meet the regulated minimum wage. Daily profit measures the net financial outcome for platforms, calculated by subtracting daily subsidies and fixed operating expenses from daily revenue. Accumulated capital reflects the cumulative net profit and loss over time.

## Impact of Platforms' Lockout Strategy

Minimum wage regulation in ride-sourcing markets can impose substantial financial burdens on platforms, potentially threatening their long-term viability. To mitigate excessive subsidy obligations, platforms typically implement a lockout strategy that restricts driver access to platform at specific times. In this study, the lockout strategy is implemented by first ranking drivers according to their participation rates in previous days, then applying a one-to-ten driver-to-traveler ratio [17] to retain only the most active drivers. Figure 4 presents the impact of varying levels of minimum wage regulation on ride-sourcing supply and demand, when platforms employ a lockout strategy—restricting driver access to the platform to prevent oversupply. Under weak regulation, the lockout strategy has a minor impact, resulting in similar supply and demand growth, with driver income stabilizing at approximately €12.50 per hour and traveler waiting time reaching 4.20 minutes. In contrast, the lockout strategy proves to be significantly advantageous for the platform under the more intense regulations. The number of active drivers under both moderate and strong regulations converges to 65, compared to 166 and 175, respectively, in the absence of the lockout strategy. This highlights that a significant number of drivers lose their jobs due to the platforms restrictive strategies. Interestingly, the lockout strategy considerably improves driver income under moderate minimum wage regulation, reaching approximately €14.20 per hour—a 15% increase compared to the same regulation without the lockout strategy. This improvement arises because, in the absence of a lockout strategy, platforms tend to lower trip fares to discourage new driver participation and thereby reduce subsidy obligations. In contrast, with a lockout strategy in place, platforms can maintain higher trip fares without triggering additional driver supply, as access is already restricted. Similarly, under strong regulation, driver income increases to €15.30 per hour, reflecting a slight improvement compared to the scenario without the lockout strategy.

Among all scenarios, weak regulation results in the highest number of travelers using the platform, reaching approximately 1070 travelers. Under both moderate and strong regulations, the number of travelers converges to 630, compared to 1322 and 819, respectively, in the absence of the lockout strategy. Across varying levels of minimum wage regulation, traveler waiting times converge to slightly different values in the presence of a lockout strategy. This indicates a lower level of service for travelers under moderate and strong regulations, with waiting times stabilizing at approximately 3.90 minutes—higher than in the absence of the lockout strategy under the same regulatory conditions.

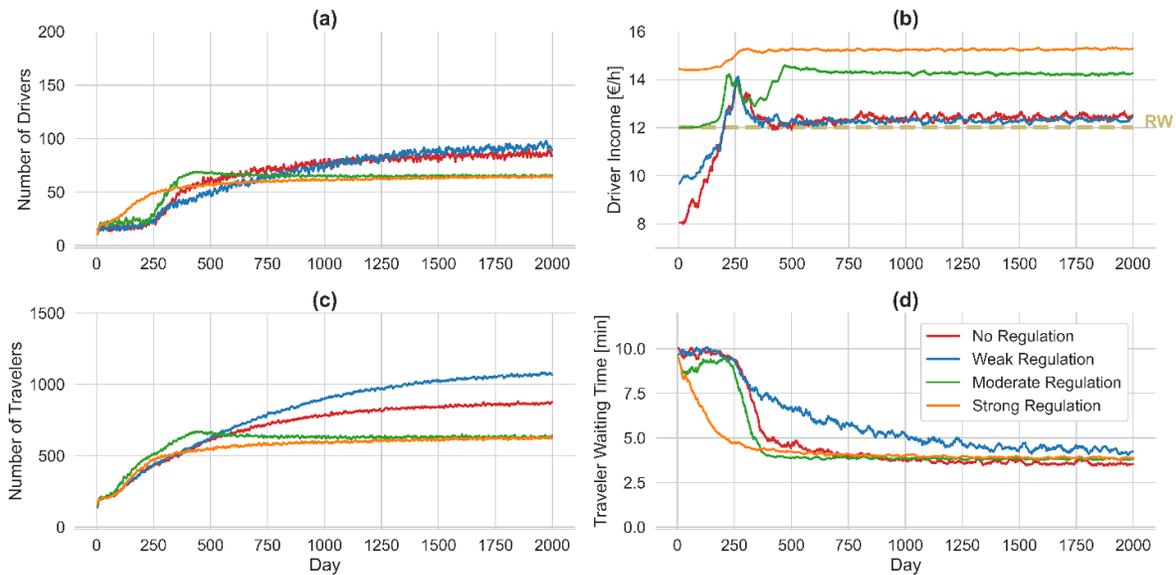

**Figure 4.** Impact of platform lockout strategy on ride-sourcing supply and demand under varying minimum wage regulations. Four scenarios are illustrated based on the presence and intensity of regulation: no regulation (red), weak regulation (blue), moderate regulation (green), and strong regulation (orange). On the supply side (top row), the number of active drivers and their average income are shown. On the demand side (bottom row), the number of travelers and their average waiting time are depicted.

Figure 5 suggests that platforms can effectively mitigate the adverse impacts of minimum wage regulation through the implementation of a lockout strategy. While weak regulation yields the highest long-term platform revenue, reaching approximately €810 per day, both moderate and strong regulations stabilize at lower levels of around €580 per day, due to a reduced number of travelers compared to scenarios without the lockout strategy. The reduction in minimum wage subsidies imposed on platforms, enabled by the implementation of the lockout strategy, demonstrates the effectiveness of this approach. In particular, platforms reduce the daily subsidy to approximately €240 and €110 under strong and moderate regulations, respectively, with the lockout strategy—compared to €2650 and €1130 per day in its absence. The low subsidy burden combined with high revenue under weak regulation enables platforms to achieve sustained long-term profitability. While platforms exhibit improved profitability under moderate and strong regulations with the lockout strategy, they still operate at a loss as indicated with negative profit. Accordingly, the accumulated losses under these two regulatory scenarios are significantly mitigated following the adoption of the lockout strategy, suggesting a higher chance of platform survival.

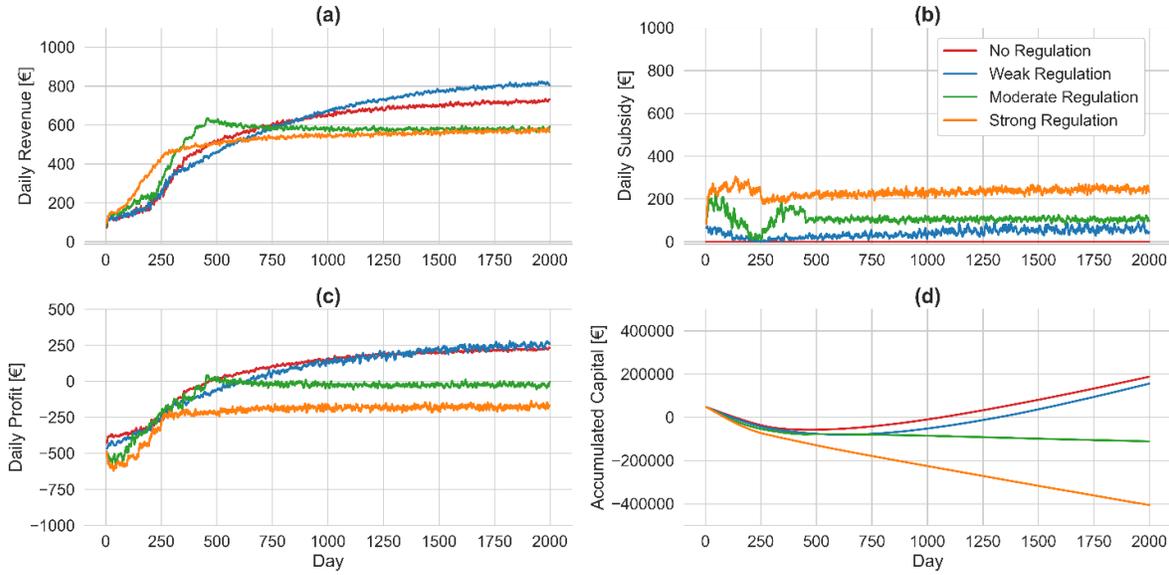

**Figure 5.** Average financial performance across two platforms under varying minimum wage regulations with lockout strategy. Four scenarios are illustrated based on the presence and intensity of regulation: no regulation (red), weak regulation (blue), moderate regulation (green), and strong regulation (orange).

We present a comparative summary of key performance indicators (KPIs) across all regulatory and lockout scenarios in Table 1. Each metric reflects the average value over the final 50 days of simulation, corresponding to the steady-state system behavior.

| | No Regulation | Weak Regulation | | Moderate Regulation | | Strong Regulation | |
|---|---|---|---|---|---|---|---|
| **Metrics** | No Lockout | No Lockout | With Lockout | No Lockout | With Lockout | No Lockout | With Lockout |
| Number of Driver | 85 | 91 | 93 | 166 | 65 | 175 | 65 |
| Driver Income [€/h] | 12.50 | 12.50 | 12.50 | 12.20 | 14.20 | 14.20 | 15.30 |
| Number of Traveler | 866 | 1051 | 1070 | 1322 | 630 | 819 | 630 |
| Waiting Time [min] | 3.50 | 4.40 | 4.20 | 2.40 | 3.90 | 2.00 | 3.90 |
| Platform Fare [€/km] | 1.40 | 1.20 | 1.20 | 1.20 | 1.40 | 1.60 | 1.60 |
| Platform Daily Revenue [€] | 721.90 | 793.60 | 810 | 1009.60 | 580 | 742.80 | 580 |
| Platform Daily Profit [€] | 221.90 | 234.10 | 251.90 | -620 | -28.80 | -2400 | -171.80 |
| Platform Daily Subsidy [€] | 0.0 | 60 | 60 | 1130 | 110 | 2650 | 240 |
| Platform Accumulated Capital [€] | 176,640 | 120,620 | 143,190 | -918,240 | -110,250 | -4,312,330 | -397,230 |

**Table 1.** Summary of results under different regulatory and lockout scenarios, averaged over the final 50 days of simulation.

## Distributions of Driver Income and Traveler Waiting Time

To gain deeper insight into individual user experiences beyond aggregate-level outcomes, we analyze the distributions of driver income and traveler waiting time. Figure 6 (a) illustrates the distribution of driver income under different levels of minimum wage regulation, both in the absence and presence of the lockout strategy. Without regulatory intervention, the distribution of driver income is relatively wide, ranging from approximately €3 to nearly €21 per hour. The distribution peaks around a mean of €12.50 per hour, with approximately 45% of drivers earning below the reservation wage of €12 per hour. We observe more concentrated distributions when minimum wage regulation is implemented, compared to the unregulated scenario, regardless of the regulation's intensity or the presence of the platforms' lockout strategy. This highlights the effectiveness of minimum wage regulations in reducing income inequality among ride-sourcing drivers.

Under weak regulation, drivers earn at least the minimum wage of €9.60 per hour, with some drivers earning up to approximately €18 per hour. While a bimodal distribution is observed in the absence of a lockout strategy, a unimodal distribution emerges when the lockout strategy is present. Notably, in both scenarios, the average income for drivers who complete rides so that no add-on is needed to match the minimum wage is €12.50 per hour. However, in the absence of a lockout strategy, a considerable portion—approximately 25%—of drivers earn below the minimum wage, with their income later compensated by platforms to meet the €9.60 per hour threshold, resulting in the formation of the second peak in the distribution. Under moderate regulation, we observe highly concentrated earnings for drivers around the minimum wage of €12 per hour, as a significant portion of drivers, roughly 82%, earn below this threshold and are subsequently compensated by the platform. In contrast, under the same regulation with a lockout strategy, drivers experience a wider range of incomes, with only 26% requiring compensation to meet the minimum wage threshold. Strong regulation results in the highest proportion of drivers requiring minimum wage compensation, at 90%, which decreases to 54% when the lockout strategy is implemented.

Figure 6 (b) presents the distribution of traveler waiting times under varying levels of minimum wage regulation, both with and without the lockout strategy. Similar to the income distribution on the supply side, the highest dispersion in traveler waiting times is observed under the no regulation scenario, with values ranging from 30 seconds to 32 minutes. An examination of the two halves of the violin plots under minimum wage regulations reveals more dispersed waiting times in the presence of the lockout strategy compared to its absence. As the intensity of minimum wage regulation increases, the distribution of waiting times becomes more concentrated around the mean in the absence of the lockout strategy.

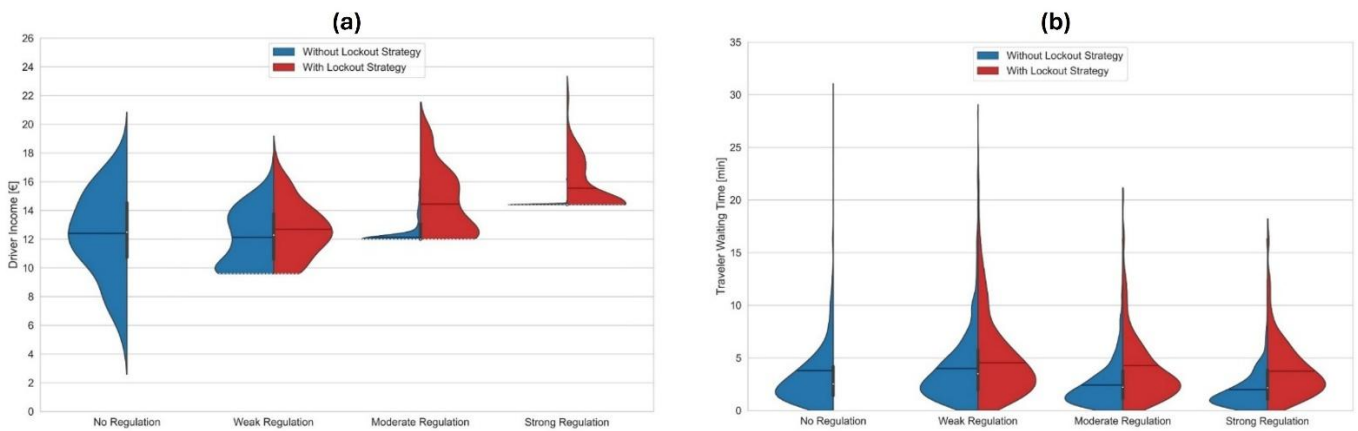

**Figure 6.** Driver income distributions (a) and Traveler waiting time distributions (b) under varying minimum wage regulations with and without lockout strategy of platforms. Each violin represents the distribution under a specific minimum wage regulation level, with the blue half indicating the absence of the lockout strategy and the red half indicating its presence. Under the no regulation scenario, the distribution is shown only without the lockout strategy, as this strategy is implemented solely in the presence of minimum wage regulation. The distributions are limited to the range of observed data, avoiding extrapolation beyond the actual values.

## Discussion

Can minimum wage regulation protect drivers without disrupting the market? This question lies at the heart of ongoing policy debates and forms the central focus of our study. While minimum wage regulation in ride-sourcing markets is primarily introduced to improve driver welfare, it has far-reaching market implications—most notably in terms of service level for travelers and the economic viability for platforms. Our findings indicate two critical factors that shape the effectiveness and consequences of this regulation. First is the implementation approach, specifically the amount of the mandated wage relative to drivers' reservation wage. Second is the platforms' response to the regulation, particularly the use of driver lockout strategies aimed at mitigating the imposed subsidy burden.

We examine varying levels of minimum wage regulation—introduced and defined within the scope of this study—both in the absence and presence of platform-imposed lockout strategies, to understand the impact of such regulation on the ride-sourcing market. Weak regulation—defined as a minimum wage set below the driver reservation wage—appears ineffective in improving driver income and has negligible impact on platform operations. As a result, the implementation of a lockout strategy under this condition proves largely redundant, with minimal observable changes in market dynamics. Furthermore, under this regulation, travelers experience a noticeable increase in waiting time, indicating a deterioration in service quality. Moderate regulation—defined as a minimum wage equal to the driver reservation wage—only benefits travelers by enhancing service level, but only in the absence of a lockout strategy. When platforms adopt a lockout strategy under this regulatory condition, they considerably reduce their subsidy obligations to drivers. Interestingly, while this response leads to improved driver income, it also results in a deterioration of service quality for travelers. Furthermore, the lockout strategy drastically reduces the number of active drivers, raising concerns that many drivers could lose their jobs as a result. Under strong regulation—defined as a minimum wage above the driver reservation wage—drivers experience the greatest improvement in earnings. While the lockout strategy substantially reduces the regulatory burden on platforms, it further enhances driver income at the expense of a significant proportion of drivers losing their jobs and a decline in service quality for travelers.

Considering that in the presence of minimum wage regulation the platforms' lockout strategy is inevitable, its implementation introduces an inherent trade-off between improving driver earnings and preserving employment opportunities in the ride-sourcing market. In practice, this implies that policy design cannot focus solely on wage levels but must also anticipate and constrain strategic platform responses. Ensuring the long-term viability of ride-sourcing services may therefore require hybrid approaches—such as conditional wage tiers linked to driver utilization and caps on deactivation or lockout durations—that both secure a fair income floor and discourage excessive rationing of labor. Only through such complementary policy measures can regulators strike the necessary balance between driver welfare, service reliability, and platform sustainability.

To the best of our knowledge, this is the first study to explore the multi-faceted impact of minimum wage regulation within a realistic environment that reproduces the market's evolutionary dynamics. Past studies lack the structural and behavioral complexity necessary to capture the dynamic, networked, and competitive nature of ride-sourcing markets. Notably, lockout strategy employed by platforms has been largely overlooked in the literature, despite its substantial implications for regulatory outcomes. We address this gap by applying an agent-based model to the city of Amsterdam, capturing interactions among key stakeholders in the ride-sourcing market. Agent-based modeling allows for detailed representation of agents—drivers, travelers, and platforms—and their adaptive behaviors in response to regulatory interventions, making it well-suited to explore the complex, dynamic nature of ride-sourcing system. Yet, the method demands extensive data and model granularity, necessitating several simplifying assumptions. Therefore, our results should not be interpreted without caveats. While each driver and traveler agent follows a unique evolutionary path, we assume a uniform reservation wage for drivers and a unform value of time for travelers, which does not account for heterogeneity. Agent behavior is restricted to day-to-day platform switching, excluding within-day multi-homing that could affect short-term market dynamics [30]. Platform competition is modeled through a discrete pricing grid, neglecting strategic levers such as commission rate or incentive schemes. Additionally, the lockout strategy is simply implemented based on supply-demand ratio across the entire simulation day.

Future research could explore more nuanced designs for minimum wage regulation, incorporating factors such as driver utilization rates to balance income improvement with job accessibility, service quality, and platform sustainability. Enhancing the modeling of platform lockout strategies—ideally grounded in empirical evidence—would also contribute to a more realistic representation of platform responses to policy interventions. Furthermore, accounting for agent heterogeneity, including reservation wage and value of time distributions, could yield richer insights into individual-level behaviors and outcomes.

## Methodology

We use agent-based modeling (ABM) to examine the impact of minimum wage regulation on ride-sourcing market dynamics, both with and without platform lockout strategies. To model the non-linear interactions between travelers, drivers, and platforms, we extend MoMaS [28], a two-sided Mobility Market Simulation framework, built on the MaaSSim simulator [31], by incorporating inter-platform competition (see Figure 7). MaaSSim simulates within-day operations by taking, as input for each simulation day, a set of participating agents (individual travelers and drivers) and platform strategies. It enables these agents to interact, and outputs both newly acquired agent experiences and platform performance metrics. This within-day modeling is essential to capture the operational complexity of ride-sourcing systems at the micro-scale, where daily market outcomes emerge from individual agent interactions and platform strategies. MoMaS uses MaaSSim outputs to update agents' participation decisions and platforms' strategies for subsequent days, enabling long-term evolution in the market. It incorporates S-shaped learning for realistic behavioral adjustment among agents and supports the implementation of various platform strategies. MoMaS is necessary for reproducing the feedback loop between short-term agent decisions and platform strategies, and the long-term implications of regulations in the ride-sourcing market. To capture the inherently competitive environment of ride-sourcing markets, we extend MoMaS with an inter-platform competition model. This model captures price wars and strategic undercutting between platforms competing to attract multi-homing travelers and drivers—an essential feature for assessing the regulatory impacts on market survival dynamics. Together, these components offer a realistic representation of within-day and day-to-day dynamics, allowing the market to evolve under both positive and negative network effects. This modeling framework is well-suited for investigating the impacts of regulatory policies in a competitive ride-sourcing environment, where travelers and drivers adapt their behavior in response to evolving market conditions and platforms strategically compete to maintain their market presence. The open-source model is available in a public repository on GitHub [32] for reproducibility.

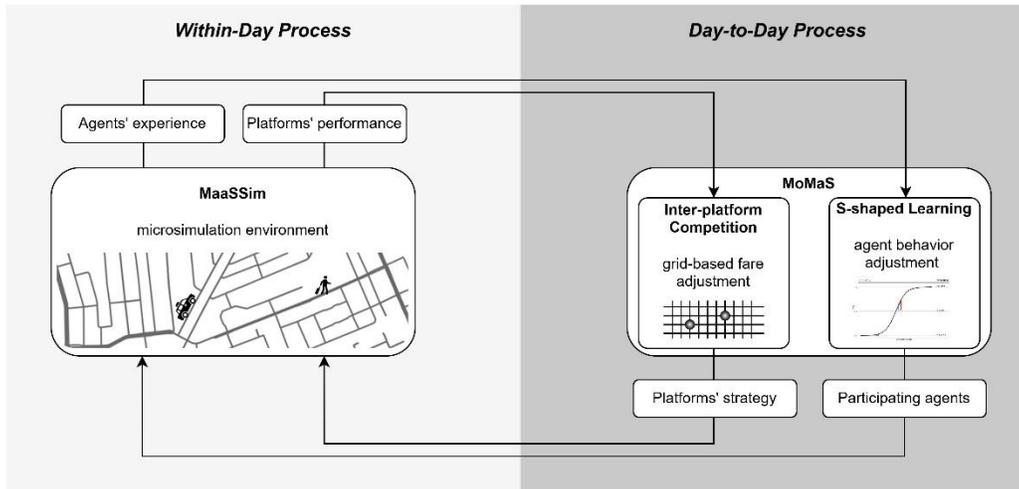

**Figure 7.** Methodology at a glance. The agent-based simulation framework, used in this study, consists of the MaaSSim and MoMaS tools. MaaSSim, as a micro-simulation environment, takes a set of agents and platform strategies, allows them to interact on a detailed road network, and yields operational outcomes. MoMaS uses the recorded interactions among stakeholders to update their actions for the subsequent day. Specifically, MoMaS relies on S-shaped learning for agents' behavioral adjustments, and we extend it by incorporating an inter-platform competition model to capture survival dynamics in the oligopolistic ride-sourcing market.

## MaaSSim

Within-day dynamics in the model are simulated using MaaSSim [31], where three types of agents interact: (i) travelers, who request rides from their origin to destination at a given time, (ii) drivers, who meet traveler demand by providing rides, and (iii) platforms, which intermediate spatio-temporal transactions between travelers and drivers. Both supply and demand are microscopic; for supply, this pertains to the explicit representation of single vehicle agents and their movements in time and space using a detailed road network graph, while for demand, this pertains to the exact trip request time and destinations defined at the graph node level. Agents are individual decision-makers. Specifically, travelers may decide which mode they use, and drivers may opt-out from the system. The matching algorithm used by platforms to link travelers with drivers is based on the myopic "first-dispatch" protocol, where travelers are paired with the closest idle driver [33].

## MoMaS

MoMaS [28] is employed as a complementary simulation framework to MaaSSim to generate day-to-day dynamics within the model. While it allows travelers and drivers to adjust their participation choice (i.e., transport mode selection) through S-shaped learning curve in response to platform strategies, platforms can implement strategies across five levers: trip fare, commission rate, discount rate, incentive rate, and marketing to maximize their revenue and market share. Each platform $a$ gradually notifies traveler and driver agents, generally denoted by $i$, about ride-sourcing option via the marketing lever. The notification status for each agent is represented with the binary variable $G_{i,t}^a$, which is set to zero and switching to one as soon as the agent is notified.

A notified agent potentially starts exploring the platform and adopts a unique day-to-day learning trajectory, supplying the demand as a driver or taking trip to its destination as a traveler. Each notified traveler $r$, on day $t$, selects among transport modes: public transport, platforms. Similarly, each notified driver $d$ chooses between occupations: working for platforms or receiving a reservation wage. By choosing reservation wage on a particular day, driver resigns from platforms and opts for the fixed wage in an alternative labor market.

The agents' participation probability is determined using a nested logit model, where an agent first chooses between the ride-sourcing nest $rs$ and other alternatives nest $o$, with $n \in N$ where $N = \{rs, o\}$, and then selects a specific option, such as a platform, from the chosen nest $a \in n$. Accordingly, as given in equation (1), the probability of choosing option $a$ is the product of the probability of choosing the nest that includes $a$ and the conditional probability of choosing $a$ given that the nest has been chosen.

$$P_{i,t}(a) = P_{i,t}(n) \cdot P_{i,t}(a|n) \tag{1}$$

The probability of choosing a nest, $P_{i,t}(n)$, depends on the expected maximum utility of all nests, $W_{i,t}^{n',p}$, each of which is calculated based on the perceived utility of options in the nest, $U_{i,t}^{a',p}$, where $\mu$ is the scale parameter at the nest level.

$$P_{i,t}(n) = \frac{\exp(\mu W_{i,t}^{n,p})}{\sum_{n' \in N} \exp(\mu W_{i,t}^{n',p})} \tag{2}$$

$$W_{i,t}^{n,p} = \frac{1}{\mu_n} \log\left(\sum_{a' \in n} \exp(\mu_n G_{i,t}^{a'} U_{i,t}^{a',p})\right) \tag{3}$$

As given in Eq. 4, the probability of choosing option $a$ from nest $n$, $P_{i,t}(a|n)$, is calculated based on the perceived utility of all choices inside the nest, where $\mu_n$ is the scale parameter within the nest.

$$P_{i,t}(a|n) = \frac{\exp(\mu_n G_{i,t}^a U_{i,t}^{a,p})}{\sum_{a' \in n} \exp(\mu_n G_{i,t}^{a'} U_{i,t}^{a',p})} \tag{4}$$

The perceived utility of agents, $U_{i,t}^{a,p}$, in MoMaS, varies within the range [0,1] and consist of three components, namely: experienced utility ($U_{i,t}^{a,e}$), marketing utility ($U_{i,t}^{a,m}$), and word-of-mouth utility ($U_{i,t}^{a,wom}$), as defined in equation (5). Experienced utility, as the primary component, is endogenous and directly derived from the simulation: drivers experience actual income, while travelers experience travel time, waiting time, and trip fare. The second component, marketing utility, is an exogenous factor reflecting the platforms' image such as marketing campaigns. Word-of-mouth, as the third component, represents perceived utility of other agents and is diffused over the social network. On a given day, when an agent encounters another agent, they exchange their perceived utility of the platform, leading to diffusion of positive and negative opinions.

The $\beta$'s in the equation reflect the relative weights of the utility components, ensuring that $\beta_i^e, \beta_i^m, \beta_i^{wom} > 0$ and $\beta_i^e + \beta_i^m + \beta_i^{wom} = 1$. The alternative-specific constant (*ASC*) captures the effect of unobserved factors on the perceived utility of alternatives and $\varepsilon_i$ is the random utility error term.

$$U_{i,t}^{a,p} = \beta_i^e U_{i,t}^{a,e} + \beta_i^m U_{i,t}^{a,m} + \beta_i^{wom} U_{i,t}^{a,wom} + ASC_i^a + \varepsilon_i \tag{5}$$

S-shaped learning model is applied to represent behavior adjustment process for the agents. Each utility component is modeled separately and updated day-to-day upon receiving a new utility signal from the respective source, such as agent's own experience, marketing campaign of platform, and peers' opinion. While the two extreme points (on lower and upper tails) of the S-shaped curve represent absolutely negative and positive perceptions, the inflection point corresponds to the neutral state. The learning process can be seen as moving along the S-shaped curve with each of the utility components. While a new positive signal pushes a component towards the upper tail, a new negative signal decreases the component towards the lower tail. In such a way, learning proceeds slowly for the agents who already have sharp, extreme opinions but rapidly for the neutral agents in response to consecutive positive or negative signals. This provides the agents with a realistic behavior adjustment (capturing reluctancy, neutrality, and loyalty) which stabilizes, and at the same time, remains sensitive to the system changes. For a detailed formulation of agents' learning, see Ghasemi and Kucharski [28].

## Inter-platform Competition Model

At the core of this study lies a novel inter-platform competition model designed to simulate adaptive fare-setting behavior between competing ride-hailing platforms. We model inter-platform competition as a turn-based pricing game on a predefined trip fare grid, as illustrated in Figure 8. The x-axis represents the trip fare per kilometer set by Platform 1, while the y-axis represents the fare set by Platform 2. Each grid point represents a specific combination of trip fares set by both platforms. Platforms begin at an initial fare point and sequentially adjust their pricing by a 0.2 [€/km] step size. For instance, if the current fare is 1.2 [€/km], a platform can choose among three options: decreasing to 1.0 [€/km], maintaining 1.2 [€/km], or increasing to 1.4 [€/km]. For Platform 1, this corresponds to moving left, staying in place, or moving right, while for Platform 2, it means moving down, staying in place, or moving up. To determine the optimal move, a platform evaluates the utility of each move by temporarily adopting the fare and recording the resulting utility over the turnover interval $\Delta t$. Indeed, we assume that a platform has perfect knowledge on the implications of their pricing adjustment strategies, which we determine by running the simulations for each possible move. As shown in equation (6), Platform 1 selects the fare $f_{p1}^*$ that maximizes its utility $U_{p1}$, given the current fare of its competitor $f_{p2}$, , and similarly for Platform 2:

$$f_{p1}^* = \underset{f'_{p1} \in \{f_{p1}-0.2, f_{p1}, f_{p1}+0.2\}}{\mathrm{argmax}} U_{p1}(f'_{p1}, f_{p2}, \Delta t) \qquad f_{p2}^* = \underset{f'_{p2} \in \{f_{p2}-0.2, f_{p2}, f_{p2}+0.2\}}{\mathrm{argmax}} U_{p2}(f_{p1}, f'_{p2}, \Delta t) \qquad (6)$$

The turnover interval, defines the duration during which only one platform can update its fare while the competitor must hold its fare constant and wait for its turn. This duration must be long enough for the effects of fare changes to propagate through the system – both supply and demand sides thereof - and affect key performance indicators such as revenue and market share. This iterative process continues until both platforms reach stable fare levels, i.e. an equilibrium state, effectively capturing the adaptive nature of price competition in the ride-sourcing market.

The utility of a specific move for platform $a$ is determined by aggregating its net gains over the turnover interval, accounting for both revenue and incurred costs. As formulated in equation (7), utility $U_a$ comprises three key components: (i) revenue, calculated as the total commissions earned from ride fares, where each commission equals the platform's share $\gamma$ of the fare $F_r$ paid by traveler for ride $r$, (ii) wage subsidy, defined as the compensation paid by the platform when a driver's income $I_{d,t'}$ falls below the minimum wage $W_{min}$, and (iii) operational expense, a fixed cost $C_a$ incurred daily to cover platform maintenance, marketing, and administrative functions. $R_{a,t'}$ and $D_{a,t'}$ denote the set of rides and drivers associated with platform $a$ on day $t'$, respectively. While $W_{min}$ and $C_a$ are predefined and constant during the simulations, other variables ($R_{a,t'}$, $D_{a,t'}$, $F_r$, and $I_{d,t'}$) are generated endogenously by the simulation as a result of the pricing decision.

$$U_a(f_{p1}, f'_{p2}, \Delta t) = \sum_{t'=t}^{t+\Delta t-1} \left[ \sum_{r \in R_{a,t'}} \gamma F_r - \sum_{d \in D_{a,t'}} \max(0, W_{min} - I_{d,t'}) - C_a \right] \qquad (7)$$

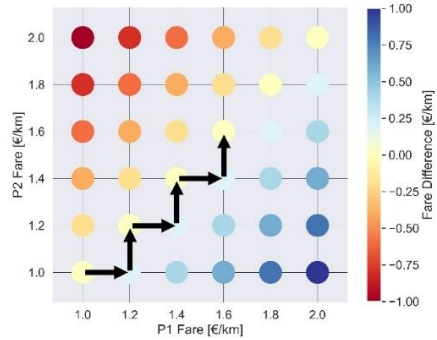

**Figure 8.** Inter-platform competition on a pricing grid. The inter-platform competition is modeled as a turn-based pricing game, where two ride-hailing platforms iteratively adjust their per-kilometer trip fares on a predefined grid. Each point on the grid represents a unique combination of fares set by both platforms, and the grid is color-coded based on the fare difference between them ($f_{p1} - f_{p2}$), highlighting competitive pricing zones. In each turn, one platform evaluates the impact of three pricing options—decreasing, maintaining, or increasing its fare—while the competitor holds its fare constant. The platform then selects the fare that maximizes its utility, which accounts for revenue from commissions, wage subsidies paid to drivers, and fixed operational costs. This process continues over successive turnover intervals until both platforms converge to stable fare levels, reflecting an equilibrium in competitive pricing.

## Minimum Wage Regulation and Lockout Strategy

Minimum wage regulation is implemented to ensure that drivers earn a minimum hourly income on their active working days, regardless of their operational activities such as the number of rides served or idle time. While drivers who already earn above the minimum wage remain unaffected, the platform is obliged to cover the difference between the earnings of lower-income drivers and the minimum wage from its own budget. The minimum wage compensation amounts to the platform's daily supplementary wage expense. The minimum wage level is assumed to remain fixed within each individual experiment.

The lockout strategy is employed by platforms as a reactive measure to mitigate the excessive expenses imposed by minimum wage regulation. Through this strategy, platforms restrict access for certain drivers, meaning that while some drivers continue to operate, others are unable to participate in service provision. As the literature and industry reports do not specify how such strategies are implemented, we assume that drivers with the highest participation rates in previous days are prioritized to stay active. In our Amsterdam case study, experiments indicate that a ratio of one active driver per ten travelers yields an efficient balance for platform growth [17]. Accordingly, in simulations with the lockout strategy, we activate the top-ranked drivers by loyalty in a one-to-ten ratio relative to the daily traveler count. For example, if 100 travelers are participating on a given day and 20 drivers would normally be active in the absence of a lockout, the platform allows only the 10 most loyal drivers to be active, thereby enforcing a one-to-ten driver-to-traveler ratio.

## Acknowledgements

This research was supported by COeXISTENCE project (grant number 101075838) which is financed by the European Research Council as well as the SUM project (grant number 101103646) which is co-funded by the European Union's Horizon Europe Innovation Action.

## Author contributions statement

F.G., A.D.R., R.K., and O.C. conceived the experiment(s), F.G. conducted the experiment(s), All authors analysed the results and reviewed the manuscript.